\documentclass[conference]{IEEEtran}
\IEEEoverridecommandlockouts
\usepackage{cite}
\usepackage{amsmath,amssymb,amsfonts}
\usepackage{algorithmic}
\usepackage{graphicx}
\usepackage{textcomp}
\usepackage{xcolor}
\def\BibTeX{{\rm B\kern-.05em{\sc i\kern-.025em b}\kern-.08em
    T\kern-.1667em\lower.7ex\hbox{E}\kern-.125emX}}
\begin{document}

\title{The Social Emotional Web\\
{\footnotesize 
\thanks{This research was sponsored, in part, by the Air Force Office for Scientific Research under contract FA9550-17-1-0327 and by DARPA (Defense Advanced Research Projects Agency) under contracts HR001121C0168 and HR00112290106}
}}

\author{\IEEEauthorblockN{Kristina Lerman}
\IEEEauthorblockA{\textit{Information Sciences Institute} \\
\textit{University of Southern California}\\
Marina del Rey, CA, USA\\
lerman@isi.edu}
}

\maketitle

\begin{abstract}
The social web has linked people on a global scale, transforming how we communicate and interact. The massive interconnectedness has created new vulnerabilities in the form of social manipulation and misinformation. As the social web matures, we are entering a new phase, where people share their private feelings and emotions. This so-called social emotional web creates new opportunities for human flourishing, but also exposes new vulnerabilities. To reap the benefits of the social emotional web, and reduce potential harms, we must anticipate how it will evolve and create policies that minimize risks.
\end{abstract}


\section{Introduction}

Arguably no other technology has transformed humanity as fundamentally and as quickly as the social web~\cite{twenge2017have}. Social media platforms like Facebook, Twitter, Instagram, Snapchat and Tiktok now connect billions of people world-wide, enabling them to share status updates, images, videos, and links to online content. The social web democratized the production and distribution of content, reducing the power of traditional gate keepers to decide what information gets attention; it created a cottage industry of influencers---ordinary people who have a gift for cultivating online audiences; it enabled myriads to stay connected to friends around the clock; it catalyzed mass protest movements; it provided a platform for a digital town square and replaced traditional forms of entertainment.

The massive interconnections, however,  created new  vulnerabilities and harms. Due to its low barrier to entry and global reach, social web has become an easy target for malicious influence campaigns and social manipulation. Foreign adversaries---notably from Russia, Iran, China---use Facebook, Twitter and other platforms to deploy armies of coordinated inauthentic accounts (automated bots, trolls, sock puppets) to inflame culture wars, incite violence, spread disinformation, and create polarization. Domestic actors also participate in influence campaigns for political and financial gain. Without traditional gatekeepers, it is harder to vet the quality of information, so misinformation and conspiracies abound. This has had profound consequences on society, undermining our trust in institutions and in democracy itself. %Online influence campaigns in some cases had real-world consequences: Russian interference in the 2016 US presidential election, inciting a massacre in Myanmar, and campaigns to spread misinformation about Covid or Russia’s role in war in Ukraine, and amplify political polarization. 

As the social web matures, we are evolving along with it. Despite the growing pains listed above, we have become more comfortable disclosing our inner thoughts and feelings online. As a result, we are getting connected to many others within the emotional web. This too will lead to a profound social transformations.  
Emotions are a fundamental part of human experience: they shape how we consume information, who we pay attention to, what we believe, and how we react to new things~\cite{van2016social}. Even without the benefit of visual cues (e.g., facial expressions), audio (e.g., tone of voice), and other physiological signals, the language of text messages conveys an emotional tone. We read other people's emotions and they  affect our own mood. Emotions spread from one person to another, synchronizing feelings of large populations, at times erupting as harassment mobs or viral memes~\cite{diresta2022how}.

Emotional connection, however, creates new vulnerabilities, and the ability to measure feelings may well enable others to better manipulate them. 
How will this social emotional web grow? How will it affect us? How will it benefit us and how may it hurt us?  To reduce potential harms and increase benefits, we must anticipate how the emotional web may evolve. In this paper, I describe the social emotional web and the opportunities and risks it creates.

\section{The Social Web: A Brief History}
The social web first captured public imagination in 2004 with sites like Del.icio.us, Digg, Flickr, and YouTube, which enabled ordinary people to share content. While some of the functionality of the new web sites was already familiar to those who used message boards and usenet newsgroups, the new generation of web sites popularized ``user-generated content'' by making it easy for anyone to create and share multi-media content in the form of web pages, news stories, photos and videos. Social web platforms  made web pages interactive, allowing users to comment on the submissions of others, upvote (like) and re-share them. The new functionality transformed how people interact with information. For example, the social bookmarking website Del.icio.us, allowed people to describe, or ``tag,'' web pages in their own words, and at the same time see what other content people have tagged with those words. Similarly, Flickr allowed people to upload photos, tag them, and also see the photos others have tagged with same words, creating an endless web of interlinked content. 

Digg pushed the envelope, merging user-generated content with the ``wisdom of crowds'' paradigm by allowing users to vote on news stories submitted by others. This way users collectively determined which new submissions should be featured on its popular front page. As another innovation, Digg allowed users to follow other people to see the new submissions they posted, thereby creating social networks that enabled some to accumulate vast amounts of attention and influence. Other social web sites, like Twitter and Facebook, extended this functionality by letting users re-share content within their social networks. 

The ability of social networks to amplify content did not go unnoticed. Wall Street Journal raved about the ``wizards of buzz''---ordinary people who have suddenly obtained power to shape what others read, watch or buy~\cite{warren2007wizards}. Today, of course, everyone is familiar with the concept of ``influencers,'' i.e., people who have the ability to promote content and make it ``go viral,'' reaching a wide audience through social connections. Predicting what content  will go ``viral,'' quantifying how the structure of social networks shaped information diffusion, and the role that influencers played in amplifying information, have become popular research topics among academics, with thousands of papers and even more citations.

\section{The Emotional Web}
The social web is poised for another transformation. Two major developments make this possible. First, although sites like Twitter and Facebook have long asked people to assess their momentary status with prompts like ``What’s happening?'' or ``What’s on your mind?'', people have grown more comfortable disclosing their private thoughts and   emotional states.  
The second major development was facilitated by advances in natural language processing. Language mediates social interactions and captures not only semantics, or meaning, of conversations but also the feelings, attitudes, and even implicit biases that cloud human judgment. 
%Second, natural language processing has matured enough to 
New computational tools enable automatic quantification of a range of emotions expressed in text, %and researchers can study emotions in text, creating new opportunities for discovery.
promising to deliver technology that understands human experience.

% Emotions are infectious. When we are online, we are bathed in a collective mood, which can affect our own via emotional contagion. Several studies have documented the infectiousness of emotions online [].
%However, while the function of the social web as the conduit for the spread of information has been well-studied, its emotional functions have received far less attention from researchers. %This is an oversight, since emotions are fundamental to human experience: they shape how people consume content, who they pay attention to, what they believe, and how they react to new information~\cite{van2016social}. Even without the benefit of visual signals (e.g., facial expressions) or audio (e.g., tone of voice), the language of text messages conveys an emotional tone. When we are online, we pick up on these emotional tones and are bathed in collective mood.

Early works on understanding emotions relied on dictionary-based methods to measure the sentiment expressed in messages by counting how many positive or negative words they contained~\cite{golder2011diurnal,bollen2011twitter}. Researchers found that the sentiment of tweets in aggregate showed the characteristic diurnal and weekly patterns of mood variation~\cite{golder2011diurnal}. Another work found that sentiment tracks the geographic distribution of subjective wellbeing~\cite{jaidka2020estimating}. A new generation of methods based on large language models enabled a wider range of emotional expressions to be quantified at scale~\cite{alhuzali2021spanemo}. Depending on training data, these models can recognize a dozen or more emotions, such as love, joy, fear, anger, disgust, etc. I illustrate with two case studies of emotions during notable recent time periods.

\begin{figure}
    \centering
    \includegraphics[width=\linewidth]{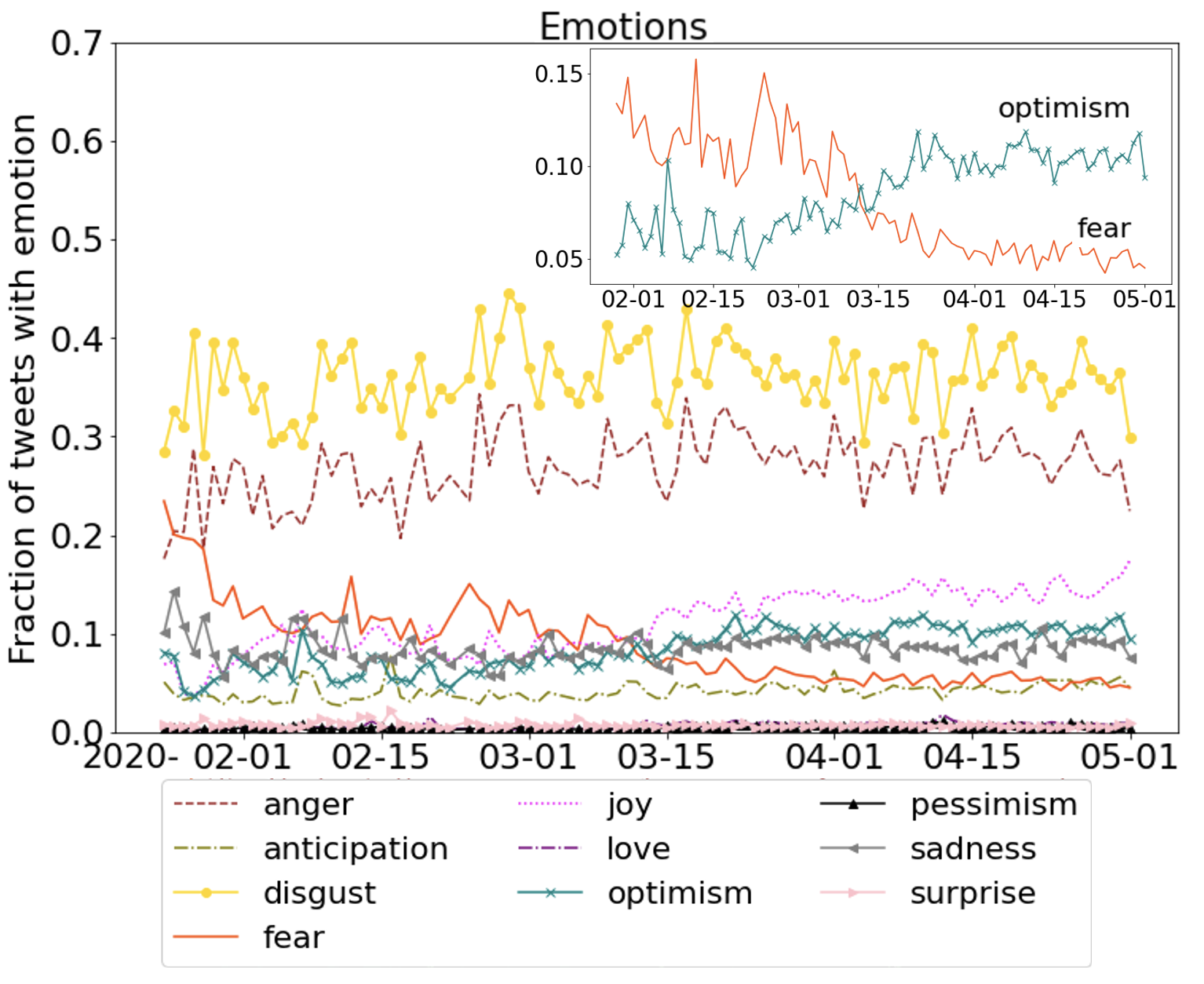}
    \caption{Trends in emotions extracted from Twitter conversations about the Covid-19 pandemic from 01/24/2020 to 05/01/2020. The plot shows the fraction of daily tweets containing the specified emotion. The inset zooms in on two emotions: fear and optimism.}
    \label{fig:timeline}
\end{figure}

\subsection{Emotions during the Covid-19 Pandemic}
In a recent study we measured dynamics of emotions during the early months of the Covid-19 pandemic~\cite{guo2022emotion}. We used state-of-the-art language models to measure emotions, such as fear, anger, and optimism. We found that disgust and anger were dominant emotions. We also found that positive emotions like joy and optimism increased starting in late February through the end of March 2020, while fear decreased. This is surprising, as this time period was punctuated by the first US death, the declaration of national emergency, and lockdowns in many municipalities. The drop in fear was accompanied by the decrease in uncertainty, as measured with a lexicon-based approach~\cite{guo2022emotion}, and an increase in expressions of solidarity. These findings suggest that people used social web to regulate their emotions during a time of crisis. 

\begin{figure*}
    \centering
    \includegraphics[width=\linewidth]{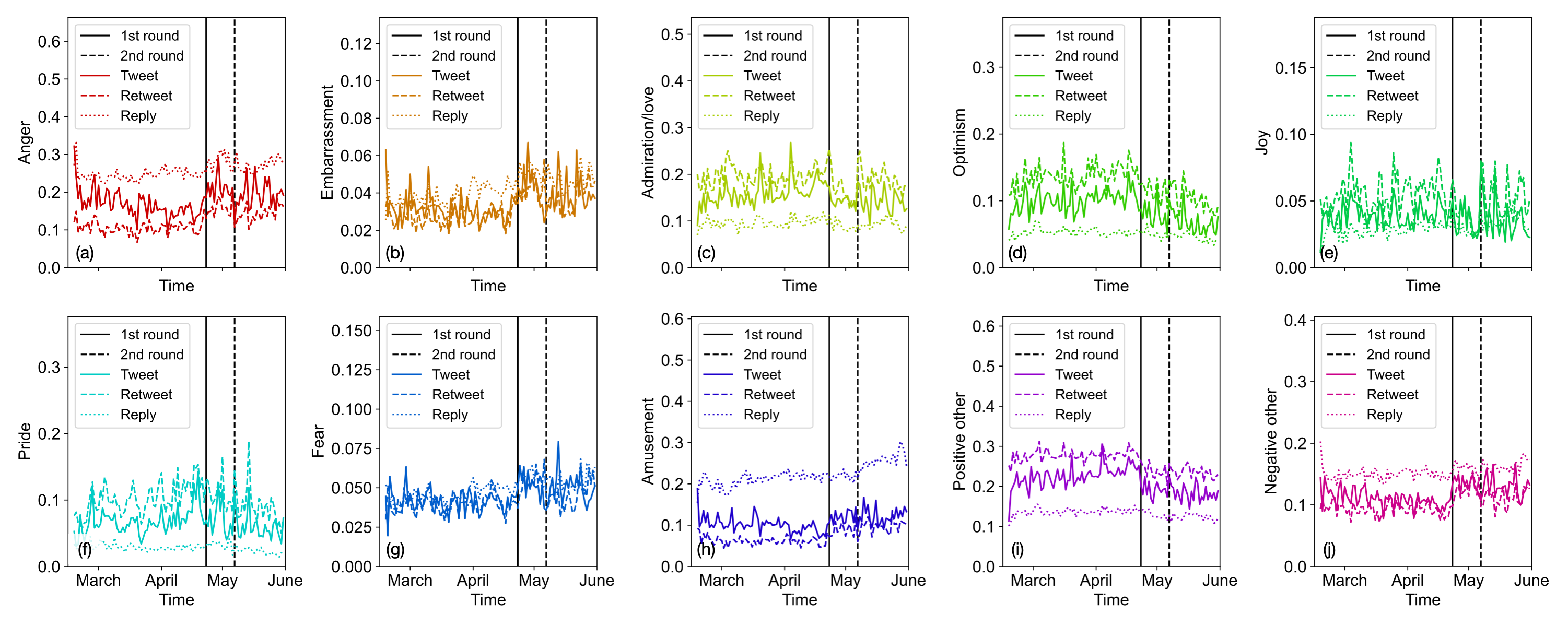}
    \caption{Trends in emotions during the 2017 French presidential elections. The plot shows the fraction of daily tweets from frequent posters that include a specific emotion: (a) anger,  (b) embarrassment,  and (c) admiration or love, (d) optimism,  (e) joy,  (f) pride,  (g) fear,  (h) amusement, (i) positive-other and (j) negative-other emotions. Solid lines represent original tweets, long dashed lines represent retweets, and short dashed lines represent replies. The lines demarcate the two rounds of voting.}
    \label{fig:timeline2}
\end{figure*}

\subsection{Emotions during the 2017 French Elections}
We analyzed a set of tweets about the 2017 French presidential election. %The tweets were collected by querying Twitter with a set of keywords like ``election'', ``élection'', ``l'élection'', ``Elysee 2017'', ``Elysee2017'', etc., and were mainly in French.
We focused on tweets from prolific users (with more than 400 posts) and automatically labeled their emotions. Figure~\ref{fig:timeline2} shows the mean daily weight of emotions in these tweets. Outside of the election period, marked by the two vertical lines, positive emotions, like admiration and optimism, were more common than negative emotions. Interestingly, pride peaked at Macron's inauguration on 14 May 2017. Emotions in replies were less common, while positive emotions were more common in retweets compared to other types of tweets. Both of these observations suggest that retweets indicate agreement and replies indicate disagreement. Negative emotions rose, and positive emotions fell, after the first round of voting.  Positive emotions were not suppressed for long: just after the second round of voting, joy, admiration, and optimism all spiked. 

These results highlight the opportunities to use emotion recognition in the study of elections and to create better tools for early detection of online influence campaigns.

\section{Opportunities and Risks}
The social emotional web connects people on an intimate level, allowing them to express their own emotions and perceive the emotions of others. Emotions are ``contagious'': they can spread from person to person even in the absence of interpersonal interactions or non-verbal cues~\cite{kramer2014experimental}. This interactivity holds a promise for societal transformation. If used properly, it can become a powerful tool for improving health and wellbeing. In the wrong hands, however, it can wreak more havoc than trolls and social bots do today. Anticipating some of the harms will help us proactively work to reduce them.

\subsection{Opportunities}

\paragraph{Social Support and Belonging}
At its best, the emotional web provides social support by linking individuals with like-minded peers.  People afflicted with rare medical conditions can seek out others with the same diagnosis to provide moral support or share information about treatments and therapeutics. LGBTQ youths have found comfort in finding role models online that they lack within their own communities. Many other special affinity groups are active online providing support, information, and community to geographically dispersed members, fulfilling the fundamental human need to belong~\cite{baumeister2011need}. This can feel especially empowering to dissidents in countries with repressive regimes who need a community to sustain dissent.

\paragraph{Identity Formation}
Individual and group identities are now largely formed online through individual expression and group interactions. Along the same lines, social media can catalyze social activism, allowing masses to quickly organize around a cause. We saw this very dramatically during the global racial justice protests kindled by George Floyd's killing. But the web has also energized extremism, globally linking dispersed and disparate extremist groups --- from misogynistic incels to white supremacists --- within the web of hate~\cite{johnson2019hidden}. 
%While it is easier than ever to bring people together in a social protest, the very ease may have undermined the long-term viability of social movements. By reducing the effort required to bring people together [Tufekci 2022]

\paragraph{Collective Sensemaking and Emotional Regulation}
Social web aids sensemaking in times of crisis, which in turn promotes resilience by helping people better regulate their emotions. This is a vital function, since disasters have proliferated on a global scale, driven in large part by climate change and fluid economic and political world order. Disasters create uncertainty: which roads have been washed out by flooding, what non-pharmaceutical interventions are effective in limiting the spread of an infectious disease, or who is the lawful election winner. By sharing information and seeking opinions from one’s social group can help reduce uncertainty and fear. This may explain the rise in optimism and a decrease in fear observed in pandemic-related tweets during March of 2020~\cite{guo2022emotion}. Although the news were grim---Americans were beginning to die from the virus and communities were locking down---it seemed possible in those early months to beat the novel coronavirus by sheltering at home (``15 days to slow the spread'' or ``two weeks to flatten the curve''). 
Whether by reducing uncertainty or %emotional contagion, 
amplifying solidarity, the social web helps people to collectively regulate their emotions, especially during crises.

\subsection{Risks}

\paragraph{Negativity Bias}
There is an inherent asymmetry in human emotions that makes ``bad stronger than good''~\cite{baumeister2001bad}. Negative emotions have stronger impact than positive emotions, criticism affects us more than praise, and negative first impression is harder to overcome.
We are primed by evolution to pay more attention to hazards, i.e., negative information, than to benefits, i.e., positive information. 
%Polarization
The negativity bias makes threats more believable and more likely to be shared~\cite{fessler2014negatively}, resulting in more attention to negative information. 
Moreover, there is ideological asymmetry whereby social conservatives are more attentive to threats than others~\cite{fessler2017political}. This may explain political asymmetry in misinformation noted by recent studies, which found that conservatives see~\cite{rao2022partisan} and share~\cite{nikolovright} more misinformation than liberal Twitter users. 

% Large disparities from small causes: feedback loops amplify
\paragraph{Psychological Contagion}

Emotions, including happiness and stress, can spread from person to person even in the absence of physical interactions, a phenomenon known as psychological contagion. Since negative emotions are more salient, they are more likely to spread by psychological contagion, lowering the collective mood. 

While cases of psychological contagion through social media have been documented~\cite{kramer2014experimental,coviello2014detecting}, many examples of contagion through traditional broadcast or print media exist. For example, before western TV programming was introduced in Fiji in the mid-1990s, eating disorders were unknown in this island nation. However, they soon rose dramatically among girls~\cite{becker2004television}. As an older example, journalists are careful to report on suicides due to the risk of copycat suicides. This phenomenon is so prevalent it has been dubbed the Werther Effect, after the 1774 Goethe novel, whose protagonist's death by suicide was imitated by many young men.

\paragraph{Feedback Loops}
Feedback loops are an important mechanism in the life cycle of complex systems: they  magnify small initial differences to create large disparities in outcomes. For example, in science, feedback loops are responsible for large disparities in researcher recognition. They  arise due to the Matthew Effect~\cite{merton1968matthew} (i.e., cumulative advantage), by which scientists who are already advantaged get even more recognition than their less-advantaged peers. Feedback loops also amplify slight biases in individual decisions to create large racial and gender disparities in education, healthcare, law enforcement, and other fields~\cite{o2016weapons}.

Feedback loops can arise on the social web due to synergies between psychological contagion and peer pressure, which signals collective approval or disapproval of a behavior.
%Social interactions can further incentivize (or disincentivize) emotions or behaviors via peer pressure effect that signals social approval or disapproval of the behavior. Together,  psychological contagion and social approval promote positive (pro-social) or negative  (anti-social) behaviors. 
Feedback loops can result in both positive (pro-social) and negative (anti-social) behaviors.
Support groups create positive feedback loops (aka virtuous cycles), where individuals model positive behaviors and receive approval for them. However, more often than not, the feedback loops create vicious cycles that promote anti-social behaviors. %they can  amplify the initially small differences in negativity to create large disparities in emotions. 
Take, for example, a person who has an angry outburst online, making others angrier through psychological contagion. If the angry posts get more community attention in the form of engagement and interaction (e.g., replies, likes), the angry behavior will persist and grow~\cite{brady2021social}. Algorithms may further compound the problem by highlighting posts with more engagement.
%This can have profound undesirable consequences on vulnerable individuals.
Online harassment, hate speech, and other unfettered negative expressions could be driven by such negative feedback loops.
However, the same mechanisms create feedback loops also make them fragile to interventions, and even small mitigation measures can have large effect on outcomes.

\paragraph{Health and Wellbeing} 
There has been a marked decline in mental health and wellbeing of children and adolescents over the past decade~\cite{twenge2017have}. Although there is still no consensus on the causes of the decline, it is clear that social media contributes to stress that erodes psychological wellbeing in multiple ways, for example, by reducing the amount of time for in-person interactions and sleep. Another mechanism for the erosion fo wellbeing is the psychological contagion of stress, which is further compounded by feedback with peer approval or disapproval,  creating a vicious cycle of mental health deterioration.  

This is especially harmful when psychological contagion is linked to a psychogenic illness, a condition in which mental stressors create physical symptoms of a disease. Psychogenic illnesses among adolescents grew rapidly during the pandemic, mediated in no small part by social media. For example, the number of teens  who exhibited Tourette-like tics (involuntary physical movements or vocalizations) skyrocketed. The similarity of tics led doctors to suspect that they ``caught'' them from online influencers, the first known case of \textit{mass psychogenic illness}~\cite{muller2022stop}. Other  psychological illnesses that increased rapidly during the pandemic, like eating disorders, could similarly be a response to the emotional stressors of the pandemic, amplified by the feedback loops of the emotional web. Vulnerable girls seeking  diet tips online may end up discovering ``pro-ana'' and ``pro-mia'' groups that promote anorexia and bulimia  by sharing tips on how to tolerate extreme hunger and cheering members who severely restrict food. By falling into these groups, the girls get trapped in a vicious cycle that leads to an eating disorder.

%Although few cases of psychological contagion through social media have been documented, many examples of media contagion exist. For example, before western TV programming was introduced on Fiji in the mid-1990s, purging as a way to control weight was unknown. After TV, purging in girls rose to 45\% by 2007, according to one study. Due to the risk of copycat suicides, responsible journalists now follow guidelines on reporting about suicides.

\paragraph{Social Networks and Social Comparisons}
The social web can also erode wellbeing through negative social comparisons. Human are wired to figure out their place within groups~\cite{fiske2011envy}. Normally, this process helps establish status hierarchies, which help groups coordinate to achieve common goals or manage common resources. However, the social web hijacks the build-in cognitive and psychological mechanisms of social comparison to deleterious ends. The algorithms that curate social media feeds present streams of viral content from influencers featuring unattainable ideals of beauty, fun and power, making the viewers feel worse about themselves.  
Even the structure of online connections can distort social comparisons~\cite{Lerman2016majority} by making  friends seem more popular, desirable, and even happier~\cite{bollen2017happiness} than they really are.
These effects fuel FOMO---the fear of missing out---and the pain of being excluded over time degrades wellbeing.

\paragraph{Manipulation and Misinformation}
Emotional connection creates economic opportunities that expose people to exploitation. In the battle for hearts and minds, the platforms hold an advantage, since they can learn your emotional DNA and then use this information for targeted advertisements.
This is already happening. Platforms like Tiktok and Instagram are learning individual psychological vulnerabilities based on their patterns of content consumption and then bait users with distressing content they cannot resist. In one example, Tiktok showed messages about pregnancy loss and still birth to pregnant women~\footnote{https://www.latimes.com/business/technology/story/2022-05-25/for-pregnant-women-the-internet-can-be-a-nightmare}, suicide content to depressed teens,\footnote{https://www.dailymail.co.uk/news/article-11268609/We-posed-TikTok-teen-suicide-posts-appeared-minutes.html} and dieting and weight-loss content to young girls.\footnote{https://www.theguardian.com/society/2021/jul/20/instagram-pushes-weight-loss-messages-to-teenagers} 

People turn to online authorities to make sense of the world. But, when truth hurts, demagogues use lies to soothe their followers' feelings. 
%Emotional appeals can stoke political polarization. Emotionally charged messages are used to rally supporters and enrage the opposition, often without regard for the factuality of the information they convey. 
As a result, misinformation has proliferated on the social web and proved surprisingly resistant to mitigation efforts, like fact checking. Facts will not eradicate misinformation, because it serves a psychological need to make people feel better. Efforts to combat misinformation should include steps to understand the psychological needs it fulfills.

\section{Conclusion}
Technology is changing us. The social web (in concert with smartphones) has already altered how we interact with others and how we spend our time. Our transformation is still in its nascent stages. The next step of the development will involve %the emotional web, which will facilitate 
creating emotional connections that enable us to share our loves and joys, but also anger, disappointment and disgust. In parallel, we are creating tools that can better recognize the emotions in text, but soon also in images and videos. % Metaverse? Dissociation from the body – effects on the brain and mental health?
This will allow us to build technologies that understand people on a deeper psychological level, that can read our emotions and adapt to our moods.

The real-time connections between our own emotions and those of billions of others will create complex systems with unpredictable trajectories, requiring new advances in the science of complexity. This may also lead to qualitatively new phenomena, like the formation of the Global Brain, a thinking, feeling super-organism that may transcend humanity. Or the emotional web may be hijacked by commercial interests.  How the social emotional web will upend our lives, whether it brings us closer together or tears us apart, is still a matter of conjecture, and imagining potential futures is the best way to bring about the positive vision.

\section*{Acknowledgment}
I am grateful to my students and colleagues for their important contributions to this new research area: Ashwin Rao, Fiona Guo, Fred Morstatter, Georgios Chochlakis, Kai Chen, Keith Burghardt, Negar Mokhberian, Rebecca Dorn, Shri Narayanan, Zihao He.

% [Jamin Warren, John Jurgensen, The Wizards of Buzz, WSJ, Feb 10, 2007]

% Kirsten R Müller-Vahl, Anna Pisarenko, Ewgeni Jakubovski, Carolin Fremer, Stop that! It’s not Tourette’s but a new type of mass sociogenic illness, Brain, Volume 145, Issue 2, February 2022, Pages 476–480, https://doi.org/10.1093/brain/awab316

% vanKleef, G. A.; Cheshin, A.; Fischer, A. H.; and Schneider, I. K.
% 2016. Editorial: The Social Nature of Emotions. Frontiers in Psychology 7: 896

% Brady, W. J., McLoughlin, K., Torres, M., Luo, K., Gendron, M., \& Crockett, M. (2022). Overperception of moral outrage in online social networks inflates beliefs about intergroup hostility.

% Brady, W. J., McLoughlin, K., Doan, T. N., \& Crockett, M. J. (2021). How social learning amplifies moral outrage expression in online social networks. Science Advances, 7(33), eabe5641.
\bibliographystyle{plain}
\bibliography{references}

% \begin{thebibliography}{00}
% \bibitem{b1} G. Eason, B. Noble, and I. N. Sneddon, ``On certain integrals of Lipschitz-Hankel type involving products of Bessel functions,'' Phil. Trans. Roy. Soc. London, vol. A247, pp. 529--551, April 1955.
% \bibitem{b2} J. Clerk Maxwell, A Treatise on Electricity and Magnetism, 3rd ed., vol. 2. Oxford: Clarendon, 1892, pp.68--73.
% \bibitem{b3} I. S. Jacobs and C. P. Bean, ``Fine particles, thin films and exchange anisotropy,'' in Magnetism, vol. III, G. T. Rado and H. Suhl, Eds. New York: Academic, 1963, pp. 271--350.
% \bibitem{b4} K. Elissa, ``Title of paper if known,'' unpublished.
% \bibitem{b5} R. Nicole, ``Title of paper with only first word capitalized,'' J. Name Stand. Abbrev., in press.
% \bibitem{b6} Y. Yorozu, M. Hirano, K. Oka, and Y. Tagawa, ``Electron spectroscopy studies on magneto-optical media and plastic substrate interface,'' IEEE Transl. J. Magn. Japan, vol. 2, pp. 740--741, August 1987 [Digests 9th Annual Conf. Magnetics Japan, p. 301, 1982].
% \bibitem{b7} M. Young, The Technical Writer's Handbook. Mill Valley, CA: University Science, 1989.
% \end{thebibliography}

\end{document}